\documentclass[traditabstract]{aa}

\usepackage{graphicx}
\usepackage{txfonts}

\newcommand{\degree}{\mbox{$^{\circ}$}}                 
\newcommand{\KS}{\mbox{$K_{\rm s}$}}

\newcommand{\iGaia}{{\it Gaia}\,\,}

\begin{document}

\title{A search for distant, pulsating red giants in the southern halo
\thanks{}}
\author{N.~Mauron\inst{1}, K.S.~Gigoyan\inst{2}, T.R.~Kendall\inst{3}, K.M. Hambleton\inst{4}}
\offprints{N.~Mauron}
\institute{Universit\'{e} de Montpellier, Laboratoire Univers et Particules de Montpellier CNRS-IN2P3,    
 Place Bataillon, 34095 Montpellier, France. 
  \email{nicolas.mauron@umontpellier.fr}
\and  NAS RA V.A. Ambartsumian Byurakan Astrophysical Observatory (BAO), Armenia
\and  18 Market Square, Northampton NN1 2DL, UK
\and Department of Astrophysics and Planetary Science, Villanova University, 800 Lancaster Avenue, Villanova, PA 19085
}

\date{Received by A\&A: ? / Accepted ??}

\abstract{To investigate the  asymptotic giant branch (AGB) population in the Galactic halo, we search for  pulsating AGB stars at a heliocentric distance $D > 50$\,kpc. Our research is based on  the Catalina Southern Survey (CSS) catalogue of variables, comprising 1286 long-period variables (LPVs) with $\delta < -20$\degree. We first focus on the 77 stars  in the cap $|b| >$ 30\degree \,for which spectral M-type or C-type classification can be derived from Hamburg-ESO objective prism spectra. Most of these  are oxygen-rich (M-type) and very few are  carbon rich.  The periods are in the range 100-500 days, and CSS amplitudes are up to 3 mag. In this small sample, no halo AGB star is fainter than
 \KS$_0 = 12.5$. This may be due  to the scarcity of AGBs  in the outer halo, or insufficient instrumental depth. Leaving aside spectral information, we then searched for even fainter pulsators ($K_{\rm s} > 12.5$) in the entire CSS catalogue. $Gaia$  astrometry makes it possible to identify some contaminants. Our final result is the identification of ten candidate distant LPVs. If these ten stars  obey  the fundamental mode K-band period--luminosity relation used for Miras and small-amplitude Miras, their distances are between 50 and 120 kpc from the Sun. In a diagram showing distance versus {\it Gaia} tangential velocity, these ten stars  have positions consistent with that of other objects in the halo, such as globular clusters  and dwarf galaxies. We  detect some underluminous AGBs that deserve further study.  Finally, the halo LPVs ressemble the slow redder variable of globular clusters when colour and periods are compared. A detailed catalogue of the 77  high-latitude M or C stars will be made available at the CDS.}

\keywords{ stars: AGB -- Galaxy: halo -- Galaxy: stellar content } 
\titlerunning{Searching cool pulsating giants in the Galactic halo} 
\authorrunning{N.~Mauron et al.}
\maketitle

\section{Introduction}
 
 The stellar populations located in the Galactic halo contain information about the general properties of the Galaxy, such as its mass, formation and evolution (Helmi \cite{helmi08}; Bland-Hawthorn \& Gerhard  \cite{blandhawthorn16}; Deason et al.\ \cite{deason20}). They are also essential for comparing the diverse halos of similar galaxies (Monachesi et al.\ \cite{monachesi16}; Harmsen et al.\ \cite{harmsen17}). Large and recent surveys use RR Lyr variable stars (Hernitschek et al.\ \cite{hernitschek16}; Sesar et al.\  \cite{sesar17}; Stringer et al.\ \cite{stringer20}), blue horizontal branch stars (Thomas et al.\ \cite{thomas18}; Starkenburg et al.\ \cite{starkenburg19}), K-type giants (Xue et al.\ \cite{xue15}; Janesh et al.\ \cite{janesh16}; Yang et al.\ \cite{yang19}), or M-type giants (Bochanski et al.\ \cite{bochanski14}). In this paper we focus on the population of evolved long-period variables (LPV).  We follow previous works by Huxor \& Grebel (\cite{huxor15})  centred on carbon stars, by Grady et al.\,(\cite{grady19}) who used LPVs as age indicators in the Galactic components and as tracers of merging debris, and Mauron et al. (\cite{mauron19a},\ \cite{mauron19b}) who studied M-type LPVs in the northern halo and Sagittarius tidal arms. 

Long-period variables evolve along the asymptotic giant branch (AGB) and deserve attention because they are more luminous than some of the tracers cited above. Miras or small-amplitude Miras are rarer than giants by more than an order of magnitude, but, like RR Lyr, follow the period--luminosity (PL) relation. There are old LPVs   in many globular clusters (GCs) with periods  in the range of $\sim$\,30--250 days. 
If the nascent Galaxy contained a large number of now completely dissolved GCs (Gnedin \& Ostriker \cite{gnedin97}; Odenkirchen et al.\ \cite{odenkirchen01}; Lee et al.\ \cite{lee04}), one might expect a fossil population of cluster LPVs in the halo. Similar and more numerous LPVs can also originate from dislocated dwarf galaxies. 

The old AGB  populations in GCs are M-type. These cluster  
stars are in most cases semi-regulars (SRa or SRb type), or, more rarely, Miras possessing a large flux amplitude 
$\Delta V> 2.5$\,mag (Clement \cite{clement01}, \cite{clement17}).  The  halo AGB population also comprises cool N-type carbon (C)  stars that
 have intermediate ages; from the census of halo C stars of Huxor \& Grebel (\cite{huxor15}),  
there are about\,40 stars at more than 50\,kpc from the Sun, with 9 at more than 100\,kpc. 
Roughly two-thirds of cool N-type C stars belong to the Sgr stream. Concerning the M stars, 
Mauron et al.\ (\cite{mauron19b}) presented a catalogue covering the section of sky from $\delta > -20$\degree. 
These latter authors found 57 stars at $D > 50$\,kpc and 6 at $ > 100$\,kpc; again, about two-thirds of them are in the Sgr stream.

Here, our goal is to increase the known sample of  LPVs with $D > 50$\,kpc by examining the complementary region  
with $\delta < -20$\degree. We use the Catalina Southern Survey (CSS) LPV catalogue of Drake et al.\,(\cite{drake17}), and 
separate the M-type from C-type stars  by exploiting  objective prism plates.  Attention was paid to 
contaminants: in  Mauron et al.\,(\cite{mauron19a}), we discovered that numerous  faint and  distant stars are   due to instrumental 
artefacts.   Thanks to the {\it Gaia} EDR3 data,  some supplementary contaminating stars  can be discarded.  After 
analysing the data in Sect.\,2, we discuss our results in Sect.\,3. We finally conclude that very few LPVs are as distant 
as 50 kpc in the CSS survey, and only ten distant candidates are found.

\section{Analysis}
The CSS catalogue contains  1286 LPVs distributed over an area 
of  $\sim$ 10000 square  degrees at a declination of $\delta < -20$\degree\,. In particular, it provides  periods $P$, time-averaged  magnitudes  $m_{\rm CSS}$, and amplitudes $\Delta m_{\rm CSS}$,  based on observations made from 2005 to 2013.  Its limiting magnitude is $m_{\rm CSS} \sim 17$. The CSS instrument has no filter, and for red stars similar to those of this work, $m_{\rm CSS}$ is close to a $I$-band magnitude. A majority of these 1286 LPVs are in the Galactic disc, with very few at high Galactic latitude, as shown in Fig.\,1 of Mauron et al.\,(\cite{mauron19a}). Also, a fraction of the halo of the Large Magellanic Cloud is covered, and a number of LPVs are  detected in  the Fornax and Sculptor dwarf galaxies located at distances of $\sim$\,140 and 90 kpc, respectively (van den Bergh \, \cite{vandenbergh00}). This shows that the CSS catalogue is a suitable resource for detecting very distant LPVs ($D > 50$\,kpc).  

Of the 1286 CSS LPVs,  93 have been spectroscopically classified  by exploiting the Hamburg-ESO objective prism survey (HES, Christlieb et al.\ \cite{christlieb04}). Of these  93 stars, 16 are false variables due to a polluting Mira (their list is in Sect.\,2).  
In Sect.\,2.1, we focus on  the remaining 77 stars and analyse their properties. In Sect.\,2.2, we proceed by considering objects too faint to be classified with HES.  Finally, in Sect.\,2.3, we extend our search over the whole CSS LPV catalogue. 
The astrometric data that we use are mostly from the {\it Gaia} EDR3 catalogue.


\subsection{M-type and C-type stars from the CSS catalogue and the  Hamburg-ESO survey}

\begin{figure}[!ht]
\begin{center}
\includegraphics*[width=7cm, angle=-90]{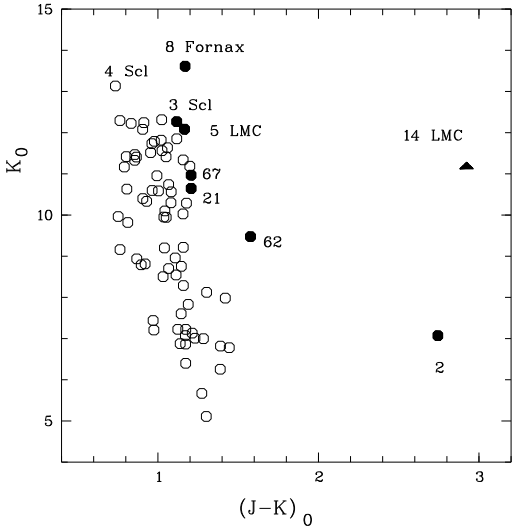}
\caption[] {Near-infrared colour--magnitude diagram of spectrally classified stars, showing the dereddened $K_{\rm s0}$ magnitude as a function of $(J-K_{\rm s})_0$ from 2MASS photometry. Uncertainties are comparable to symbol sizes.
Red objects are to the right, faint objects at the top. 
Circles: M-type stars. Filled symbols: carbon stars.  
 The labels correspond to the KG number: KG-2, 21, 62, 67 are in the halo.
Objects labelled `3 Scl' and `4 Scl' are in Sculptor. The spectrum of KG-14 (filled triangle)  is unclear, but its colour favors a carbon type.}
\label{fig1}
\end{center}
\end{figure}

Figure\,\ref{fig1} shows a colour--magnitude diagram for stars that could be classified M or C on HES plates. The interstellar  $E_{B-V}$ colour excess from Schlafly et al.\,(\cite{schlafly11}) is used to take into account Galactic extinction\footnote{The colour excess is provided by the NASA/IPAC Galactic Dust Reddening and Extinction service at https://irsa.ipac.caltech.edu/applications/DUST}. More precisely, we adopt $A_V$\,$=$\,$3.10$\,$E_{B-V}$, $A_J = 0.87$\,$ E_{B-V}$, and $A_{K{\rm s}} = 0.35$\,$ E_{B-V}$. We consider that all extincting dust is located between the Sun and the sample star, because most of the LPVs are much higher above the Galactic plane  than the thickness ($\sim$~100~pc) of the Galactic dust layer. 
It can be seen that M-type stars  largely dominate our sample.  Stars belonging to the Large Magellanic Cloud (LMC), Fornax, or Sculptor are indicated. For one object, KG-14 in the LMC, it is not clear  whether the spectrum is C or M-type, but its colour favours a carbon type that we adopt here. One can see  that a small number of carbon-rich LPV stars are detected over the imprint of the HES plates. 
Table\,\ref{TABLE1} lists their 2MASS identifications and main properties. 
 
\begin{table}
\caption[]{Properties of the stars labelled in Fig.\,\ref{fig1}. We list the KG number,
 2MASS identification, Catalina CSS-band mean magnitude, period $P$ (in days), \KS\, magnitude, 
spectral classification and membership. The spectral type of KG-14 is uncertain on HES plates, but its colour favors a C type, adopted here.
 The last column gives membership.}
\label{TABLE1}
\begin{tabular}{rrrrrrr}
\noalign{\smallskip}
\hline
\hline
\noalign{\smallskip}

KG   & 2MASS name~~~~      &    $m_{\rm CSS}$  &  $P$ &   $K_{\rm s}$ & Sp.&   \\

\noalign{\smallskip}
\hline
\noalign{\smallskip}
\noalign{\smallskip}

2    & 00165576-4400406  & 14.71 & 440  &  7.07 &  C & halo\\
3    & 00595893-3328351  & 16.23 & 196  & 12.27 &  C & Scl\\
4    & 01012084-3353047  & 16.18 &  91: & 13.14 &  M & Scl\\
5    & 01015347-6518233  & 15.71 & 162  & 12.09 &  C & LMC\\
8    & 02391532-3415083  & 17.71 & 196  & 13.62 &  C & For\\
14   & 04360427-6347499  & 17.76 & 324  & 11.21 &  C & LMC\\
21   & 04563115-3129327  & 14.26 & 147  & 10.65 &  C & halo\\
62   & 21271642-3051573  & 14.29 & 233  &  9.49 &  C & halo\\
67   & 21434114-3414311  & 14.78 & 158  & 10.98 &  C & halo\\

\noalign{\smallskip}
\noalign{\smallskip}
\hline
\noalign{\smallskip}
\end{tabular}
Note: the coordinates of KG-62  are erroneous in Mauron et al.\,(\cite{mauron19a}).
\end{table}

\begin{figure}[!ht]
 \begin{center}
 \includegraphics*[width=6cm, angle=-90]{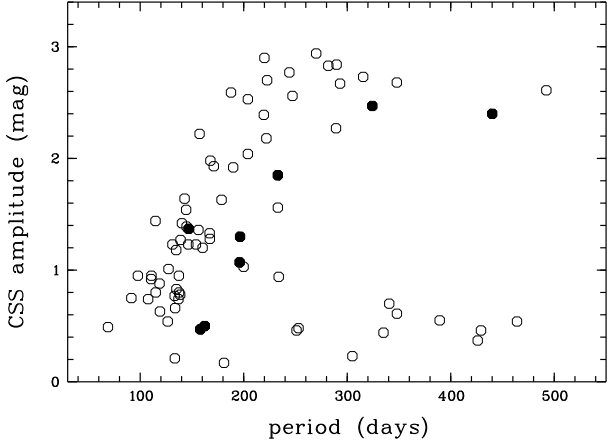}
 \caption[] { Catalina amplitude $\Delta m_{\rm CSS}$ versus period.
 Circles: M-type stars. Filled circles: carbon stars. The two stars in the upper right-hand corner are the halo carbon star KG-2 ($K_{\rm s} = 7.1$, $D \sim 14$~kpc, discussed in the text) and the M-type star \object{WOH~S24} (alias KG-15; $K_{\rm s} = 10.0$ at 51~kpc), located in the Large Magellanic Cloud.} 
\label{fig2-PA}
 \end{center}
 \end{figure}

 In Fig.\,\ref{fig1}, faint stars (\KS $> 8$) appear slightly bluer than brighter stars, indicating that our sample is not homogeneous. This inhomogeneity is reinforced by the diagram of Fig.\,\ref{fig2-PA}, where CSS amplitude is plotted versus period. There is first a main group with period from $\sim$ 100 to 300 days, and amplitude $\Delta m_{\rm CSS}$ rising from $0.6$ to $\sim$ 3 mag. Most of the C stars are slightly to the right of this group. A second family of 10 LPVs exist with $P >220$ days and relatively smaller amplitude ($\Delta m_{\rm CSS} < 0.8$ mag) compared to objects of the main group. 
We show later that this large-period small-amplitude group is also separated with a  colour based on $Gaia$ and 2MASS photometry. Below, we
first comment on the   C stars, and discuss later on  the more numerous M stars.
 
\subsubsection{C stars}
The brighter C-type star, KG-2, has a 2MASS single-epoch magnitude \KS~$ = 7.07$, a period of $P = 440$\,days, and peak-to-peak $m_{\rm CSS}$-band amplitude $\Delta m_{\rm CSS} = 2.5$\,mag.   This star is HG-2 in the catalogue of Huxor and Grebel (\cite{huxor15}).  Neither the origin nor the metallicity is known for this star located at high Galactic latitude ($b = -71$\degree). If this star is similar to solar-metallicity, cool C stars of the solar neighbourhood, with a {\it Gaia}-based distance, then $M_{\rm Ks} \approx -8.5 \pm 0.8$ from Fig.~11 of Abia et al. (\cite{abia19}). If it is similar to carbon stars of the more metal-poor Small Magellanic Cloud (SMC), we obtain a very similar estimate, $M_{\rm Ks} \approx -8.60$ by applying 
the period--luminosity 
relation\footnote{Soszy\'{n}ski et al.\ (\cite{soszynski07}) provides for SMC C-stars of sequence C the following relation $M_{\rm Ks} = -7.15 - 4.22$\, $log_{10}(P/200)$, where $P$ is in days, $K_{\rm s}$ is the time-averaged magnitude determined using $I$-band light curves (Sect.\,3 of their paper), and where a SMC distance modulus of 18.99 is used. We note that, throughout this paper, we assume all  stars (M-type or C-type) to lie on the pulsation sequence C. Their distances would increase by a factor 1.8 if they were to lie on the more luminous  sequence C'. For more details, see Table 2 and Figs.\,3 and 4 of Soszy\'{n}ski et al.\,(\cite{soszynski07})}.

Consequently, we find the  distance of KG-2 from the Sun and  from the Galactic plane to be  $D$\,$=$\,$14$~kpc. 
Concerning the uncertainty on $D$, we have  $D = 10$**${0.2(K_{\rm s} -  M_{Ks} +5)}$ and errors on $M_{Ks}$ and $K_{\rm s}$ have to be considered. The first is typically $\sigma_{\rm M_{Ks}} = 0.15$ mag (i.e. the dispersion of the PL relation; Whitelock et al.\ (\cite{whitelock09}), their Table 2). The second originates from the difference between the required but unknown time-average K$_{\rm s}$-band flux and the available single-epoch 2MASS photometry. This difference is $\delta K_{\rm s} = 0.2 \Delta m_{\rm CSS} + 0.08$ mag (see details in Mauron et al.\, \cite{mauron19b}). Adding these errors in quadrature leads to a relative error on $D$ of 9\% for 
$\Delta m_{\rm CSS} = 0.2$, to 30\% for $\Delta m_{\rm CSS} = 2.5$.
Therefore, the distance of KG-2 is $14 \pm 4$\,kpc. This distance is larger than the one derived by Huxor \& Grebel (11$\pm 4$) because they use the Wesenheit indices and consequently take into account circumstellar extinction better  for this very red star.
KG-2  is angularly located far from the lane of C stars in the Sgr arms and its  origin is unclear. One possibility is that it may have been ejected from the Galactic disc, as found by Grady et al.\ (\cite{grady19}) for some Miras. 

 The C stars KG-21, KG-62, 
 and KG-67 are warm, that is $(m_{\rm CSS}-K_{\rm s})_{0}$\,$\sim$\,3.5-4.8, compared to KG-2 with $(m_{\rm CSS}-K_{\rm s})_{0} \sim$\,7.6. They are short-period variables (see Table\,\ref{TABLE1}) and have  $K_{\rm s}$\,$\approx$\,9-11, which is fainter than  KG-2. While KG-62 is named  HG-108 in the census of Huxor and Grebel (\cite{huxor15}), KG-21 and KG-67 are not in the list of these latter authors.   If they are members of the Sgr tidal arms, their [Fe/H] is probably between -1.2 to -0.5, which is the range of metallicity in the Sgr arms (see Fig.\,2 of Carlin et al.\ \cite{carlin18}). Because the SMC has a mean average abundance  [Fe/H] $\sim -0.73$ (van den Bergh \cite{vandenbergh00}) or $\sim -1.0$ (McConnachie \cite{mcconnachie12}), our best choice is to adopt again the SMC C-type period luminosity relation (see footnote above). We find that KG-21, KG-62, and KG-67 are at distances of 28\,$\pm$\,9, 24\,$\pm$\,5, and 35\,$\pm$\,4 kpc, respectively, with $Z =$\, 17, 17, and 26\,kpc,  respectively.  The uncertainties on $D$ essentially originate from $\Delta m_{\rm CSS} =$\,2.7, 1.6, and 0.5 mag, for KG-21, KG-62, and KG-67, respectively. 

 The star KG-8 is  located in Fornax and we classified the object as C-type. This star is listed as not-periodic in the near-infrared monitoring program by Whitelock et al.\ (\cite{whitelock09}). In their Table\,3, this star is F25006. However, a period of 196 days is clearly seen in the better quality CSS data\footnote{CSS lightcurves can be seen on the Catalina site    http://nunuku.caltech.edu/cgi-bin/getcssconedb$_{\rm -}$release$_{\rm -}$img.cgi}. Its amplitude is $\Delta m_{\rm CSS} =$\,1.3 mag. The SMC PL relation for C stars leads to a distance of 140\,$\pm$\,$25$\,kpc, which agrees with $D$(Fornax)$ = 147 \pm 12$\,kpc given by McConnachie (\cite{mcconnachie12}). Agreement is also found for  KG-3 located in Sculptor and classified C-type. With $K_{\rm s} =12.27$, $P = 196$ days, and $\Delta m_{\rm CSS} =$\,1.2, our method yields $D$\,$=$\,$75\pm 13$\,kpc, which is compatible with  $D$(Sculptor)\,$= 86 \pm 6$\,kpc from McConnachie (\cite{mcconnachie12}). 

 These agreements on external galaxies supports the view that the distances of KG-2, KG-21, KG-62, and KG-67  estimated above are plausible. This means that they are largely out of the Galactic disc, and thus we consider them to be in the halo. Our distances imply heights from the Galactic plane  of between 10 and 26 kpc, much larger than 1 kpc seen for cool C stars of the disc (Abia et al.\ \cite{abia19}). 

\subsubsection{M stars}

The list of the 16  LPVs that could be classified M-type, but are in fact false variables, is given in Table\,\ref{TABLE2FALSE}. They are in general field dwarfs with significant proper motion, angularly located close to a polluting bright Mira.

\begin{table}
\caption[]{ Sixteen false LPV objects. The columns are the KG name, 2MASS coordinates (J2000) in degrees, polluting Mira star, 
separation angle in arcsec between the CSS object and polluant, period of the KG object in days (from CSS), period P of the  Mira (from GCVS; for SS Col, 
period is from the AAVSO VSX catalog). }
\label{TABLE2FALSE}
\begin{center}
\begin{tabular}{rrrrrrr}
\noalign{\smallskip}
\hline
\hline
\noalign{\smallskip}

Star   & $\alpha$~~~~   & $\delta$~~~~  & Mira   & $\theta$ & $P_{\rm KG}$ & $P_{\rm Mira}$\\
\noalign{\smallskip}
\hline

KG-22  &  85.43000  & -38.9572  & SS Col & 26    & 364  &  358\\
KG-36  & 192.93978  & -26.7755  & EP Hya & 23    & 168  &  168\\
KG-42  & 207.26013  & -28.5105  & W Hya  & 514   & 400  &  389\\
KG-41  & 207.25596  & -28.4919  & W Hya  & 447   & 397  &  389\\
KG-44  & 209.18330  & -25.5373  & FT Hya & 32    & 214  &  216\\
KG-45  & 209.18796  & -25.5651  & FT Hya & 72    & 214  &  216\\
KG-46  & 209.43125  & -31.1054  & TW Cen & 128   & 267  &  273\\
KG-47  & 209.43687  & -31.0558  & TW Cen & 54    & 267  &  273\\
KG-48  & 209.77937  & -25.8551  & FQ Hya & 28    & 179  &  177\\
KG-52  & 219.36920  & -20.3208  & LY Lib & 28    & 289  &  287\\
KG-55  & 302.31581  & -57.3415  & BR Pav & 36    & 251  &  250\\
KG-56  & 302.31626  & -57.3173  & BR Pav & 52    & 252  &  250\\
KG-63  & 322.63793  & -53.9635  & X Ind  & 51    & 225  &  224\\
KG-83  & 340.79609  & -41.5202  & DS Gru & 46    & 255  &  260\\  
KG-89  & 352.03795  & -47.4441  & RU Phe & 49    & 291  &  288\\ 
KG-91  & 359.12315  & -49.7759  & R Phe  & 44    & 267  &  263\\  

\noalign{\smallskip}
\noalign{\smallskip}
\hline
\end{tabular}
\end{center}
\end{table}


 Among the remaining 77 stars, there are five M-type stars that are angularly close or in the LMC. We checked that their periods and \KS \, magnitudes provide a correct distance when the Soszy\'{n}ski et al. PL relation for M-type LMC AGBs is applied. These stars are KG-13, KG-15, KG-16, KG-17, and KG-18, and  we obtain 50\,$\pm$\,14, 58\,$\pm$\,19, 53\,$\pm$\,7, 51\,$\pm$\,13, and 54\,$\pm$\,10 kpc, respectively, in good agreement with the LMC distance of 51 kpc (adopting a distance modulus of $18.52 \pm 0.09$, from McConnachie\ \cite{mcconnachie12}). 

\begin{figure}[!ht]
\begin{center}
\includegraphics*[width=7cm, angle=-90]{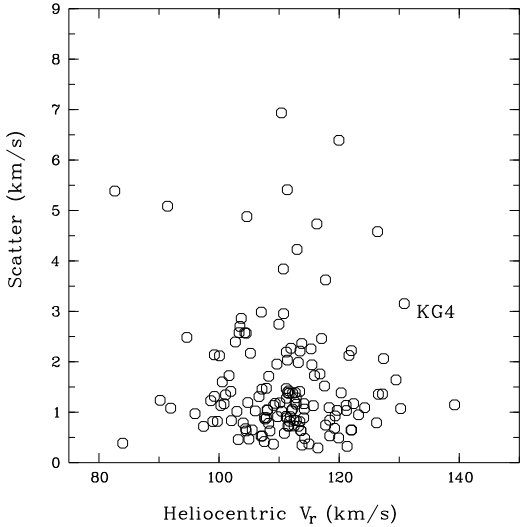}
\caption[] {  APOGEE radial velocities of stars located within one degree from the centre of the Sculptor dwarf galaxy. The LPV KG-4 is indicated.
Each point represents the average  of  11 individual APOGEE visits. This average is plotted along the X axis, and the scatter around this average is on Y axis. The median error of an individual visit is only 0.18 km s$^{-1}$. The high scatter for KG-4 is consistent with its  variability.}
\label{fig3-KG4}
\end{center}
\end{figure}

The star KG-4 is located at 0.345 deg from the centre of the Sculptor dwarf galaxy, which is 1.8 times the half-light radius of this galaxy ($r_{h} = 0.19$\,deg, from McConnachie \cite{mcconnachie12}). This star is M-type according to our examination of HES plates, and the Catalina database suggests $P$\,$=$\,91\,d, $\Delta m_{\rm CSS}$\,$=$\,0.5\,mag. However, the ASAS-SN survey (Shappee et al.\, \cite{shappee14}; Jayasinghe et al.\, \cite{jayasinghe18}) offers a slightly better light curve  and a comparable time span, and finds KG-4 to be aperiodic.  
If we apply the SMC PL relation for M-type stars as explained below, we find  $D=65 \pm 7$\,kpc, smaller than, but marginally compatible with the distance of Sculptor $86 \pm 6$ (McConnachie \cite{mcconnachie12}). Its radial velocity determined by the APOGEE experiment (J\"{o}nsson et al.\ \cite{jonsson20}) shows that KG-4 is member of Sculptor, and the important velocity scatter of individual visits is consistent with its pulsating variability (Fig.\,\ref{fig3-KG4}). 

The  next finding of this section is the identification of five M-type stars (KG-19, KG-23, KG-72, KG-88, and KG-93)  that are faint ($m_{\rm CSS} \sim 15$), with $P$ in the range 140-240 days and  $\Delta m_{\rm CSS} > 0.8$\,mag.   Their CSS names are  \object{ CRTS J044714.7-622118}, \object{CRTS J054202.5-553828},
\object{CRTS J220821.6-291754}, \object{CRTS J230418.1-405245}, and \object{CRTS J235855.3-430054}, respectively.
 We obtain their distances by adopting the  SMC period--luminosity relation for M-type stars from  Soszy\'{n}ski et al.\,(\cite{soszynski07}):\\

$M_{\rm Ks} = -7.34 - 4.14$\, $log_{10}(P/200).$ \hspace{2cm} (1)\\

 Similarly to C stars above, this SMC relation is chosen because the mean SMC metallicity ([Fe/H] $= -0.75$ to $-1.0$) coincides with the average metallicity distribution of the Sgr tidal arms and because many of our stars can be members of these arms. Therefore, we apply relation (1), which implies distances of 50 to 64 kpc. If we had adopted a LMC PL relation, the distances would be systematically (but only)  10\% smaller. 
The main characteristics of these five stars are in Table\,\ref{TABLE3}.

 The third result of this section is  that among the LPVs that could be spectrally classified, there are none fainter  
than \KS$_0 = 12.5$. We did find fainter stars that could be seen and classified M, but their 
variability is spurious (see Mauron et al.\  \cite{mauron19a} for details), or they are members of dwarf galaxies as indicated.
This lack of faint halo LPVs could be due  to an instrumental limitation. For example, identifying faint LPVs may have 
been difficult in the process of making the CSS catalogue of Drake et al.\ (\cite{drake17}) that forms the basis of our study. Also, 
spectral classification is difficult or impossible for faint stars. Alternatively, it could be due to a strong decrease in the LPV number 
with increasing distance beyond $\sim 30$\,kpc. Consequently, in the following section, we focus on the CSS 
LPVs that lie in the HES area, but could {\it not} be classified C or M.

\begin{table*}[!ht]
\caption[]{Properties of our candidate, distant CSS LPV stars (see text for their selection). $m_{\rm CSS}$ is the CSS 
time-average magnitude. $P$ is the period in days. $\Delta m_{\rm CSS}$ is the  CSS amplitude (peak to peak). $K_{\rm s}$ and $J-K_{\rm s}$ are from 2MASS. 
$A_{\rm V}$ is the Galactic V-band extinction in mag, and $A_{K}/A_{V} = 0.11$.
 $D_{\rm PL}$ is the heliocentric distance in kpc derived with the PL sequence C relation for M-type (Equation 1). 
$\delta D$, in kpc, is the distance uncertainty (see text for details). $\mu$ is the {\it Gaia} EDR3 proper motion (see also Table 4).
$V_{\rm t}$ is the tangential velocity in km s$^{-1}$ derived from  $D_{\rm PL}$ and $\mu$. The KG stars are M-type. 
The spectral types of stars ABCDE are unknown.}
\label{TABLE3}
\begin{center}
\begin{tabular}{lrrrrrrrrrrrrrr}

\noalign{\smallskip}
\hline
\hline
\noalign{\smallskip}
name    & $\alpha$~~~~ & $\delta$~~~~ & $l$ & $b$ & $m_{\rm CSS}$ &  $P$ & $\Delta m_{\rm CSS}$ &$K_{\rm s}$~~ & $J-K_{\rm s}$ & $A_{\rm V}$ & $D_{\rm PL}$& $\delta D_{\rm PL}$  & $\mu$ & $V_{\rm t}$\\
\noalign{\smallskip}
\hline
\noalign{\smallskip}
\noalign{\smallskip}

KG-19 &  71.811385 & $-$62.355530 & 272 & $-$38 & 15.13  &  139 & 0.78  & 11.82 & 1.03  & 0.10 &  50 & 7  & 1.85 & 440\\
KG-23 &  85.510835 & $-$55.641182 & 263 & $-$32 & 15.65  & 167  & 1.28  & 11.65 & 1.08  & 0.15 &  54 & 10 & 1.51 & 390\\
KG-72 & 332.090176 & $-$29.298389 & 19  & $-$54 & 15.34  & 160 &  1.20  & 11.79 & 0.99  & 0.05 &  56 & 10 & 1.00 & 240\\
KG-88 & 346.075719 & $-$40.879162 & 354 & $-$64 & 14.95  & 141 &  1.27  & 12.29 & 0.77  & 0.03 &  63 & 12 & 1.26 & 380\\
KG-93 & 359.730982 & $-$43.014912 & 333 & $-$71 & 15.15  & 235 &  1.56  & 11.41 & 0.87  & 0.03 &  64 & 14 & 0.75 & 230\\

\noalign{\smallskip}
\noalign{\smallskip}

A       & 76.208748  & $-$40.221203 & 244 & $-$37   & 17.44 &  108 & 0.55 & 14.16 & 1.06 & 0.08 & 119  & 14 & 0.26 & 150\\
C       & 141.992615 & $-$34.560715 & 262 & $+$12   & 17.33 &  139 & 1.30 & 13.61 & 1.20 & 0.53 & 112  & 21 & 0.07: & 40:\\
D       & 154.438690 & $-$29.276445 & 267 & $+$23   & 16.36 &  184 & 0.45 & 12.68 & 1.11 & 0.20 &  93  & 10 & 0.29 & 130\\
B       & 208.101207 & $-$31.747896 & 318 & $+$29   & 16.92 &  128 & 1.50 & 13.64 & 1.02 & 0.18 & 108  & 22 & 0.55 & 280\\
E       & 217.840607 & $-$36.831627 & 324 & $+$22   & 15.76 &  110 & 0.80 & 12.81 & 0.88 & 0.26 &  64  &  9 & 0.72 & 220\\

\noalign{\smallskip}
\noalign{\smallskip}
\hline
\end{tabular}
\end{center}
\end{table*}

\subsection{Stars without M/C classification from the HES}

There are 35 LPVs that could not be classified because no spectrum is visible on the 
Hamburg plates or because of overlapping spectra. Their Catalina $m_{\rm CSS}$ magnitudes range from 15
to 18. Of these 35, 7 are in Fornax, 19 can be associated with the Large Magellanic Cloud, and
6 are false faint Miras (because their light is contaminated by a true nearby, bright Mira). 

Among the remaining 3 unclassified LPVs, the first (hereafter named A) 
has remained unnoticed in the literature so far, but is interestingly faint ($K_s=14.2$). The 
second (called B) was mentioned by Grady et al.\,(\cite{grady19}) among their six 
most distant stars, and could be at $\sim$ 100 kpc. These two stars are discussed below. 
 The CSS names of A and B are \object{CRTS J050449.9-401314} and \object{CRTS J135224.2-314452}.                     
Their properties are given in Table\,\ref{TABLE3}.
Finally, the third star is a confirmed halo C-type star (2MASS\,J22065366-2506282)
with  $m_{\rm CSS} =14.07$, $K_{\rm s}=8.95$, $P= 328$\,days, and named HG-111 by Huxor \& Grebel (\cite{huxor15}). 
It is possible that this  star escaped our classification because  a deep minimum occurred 
when the HES plate was exposed.  

To conclude, when considering the 35 LPVs lying in the area of the Hamburg survey but 
not  classified M or C, we find only two interesting LPVs, stars A and B listed in Table\,\ref{TABLE3}. 

\subsection{All other stars from the CSS catalogue}

In the preceding sections, the stars A and B are faint ($K_s =$ 13.6 and 14.2) 
and presumably distant, at about 100\,kpc if the PL relation applies. Here, we present the extension of 
our search to all CSS LPVs, and not only those in the HES imprint. We first noticed
 one peculiar object, 2MASS\,J15204636-2533017, with  $l=340$\degree, $b=+26$\degree, 
$m_{\rm CSS} =14.93$, $K_{\rm s}=11.70$, $J-K_{\rm s}=1.01$. Its period is 323\,days, and
$D \sim$ 90\,kpc if the $K_{\rm s}$-band PL relation is valid ($M_{\rm Ks} = -8.2$). 
However, a strong indication that it is closer to us is 
its proper motion of 1.4 $\pm 0.1$ mas/yr, which implies a relatively high tangential 
velocity ($V_{\rm t}$) of 640 km/s if the large distance is 
correct\footnote{$V_{\rm t}$(km/s) $= 4.75 D$(kpc)$ \times \mu$(mas/yr)}. 
In addition, the light-curve amplitude is rather small, $\Delta m_{\rm CSS} = 0.32$\, mag. 
Therefore,  
this star is likely  not a pulsating AGB. It might be 
an ellipsoidal red giant variable similar to those of the LMC, which are $\sim$\,2.5
mag less luminous than AGB stars (see Fig.\,2 of Soszy\'{n}ski et al.\, \cite{soszynski07}). 

To find very distant LPVs, we  selected objects with $K_{\rm s0}$\,$>$\,$12.5$ and 
obeying $(m_{\rm CSS}-K_{\rm s}) > 2.0$ as our 77 sample stars do.     
We found 24 stars: eight are false variable stars, 
one is in Sculptor, and nine are in Fornax. Of the six remaining stars, one has a 
proper motion of 4.6 mas/yr (discussed below). 
The remaining five include stars A and B, as expected. The three others are  
referred to as C, D, and E below.  The CSS names of stars C, D, and E are \object{CRTS J092758.1-343339}, \object{CRTS J101745.2-291635}, 
and \object{CRTS J143121.7-364954}, respectively.
All five stars have proper motions of less than 0.8 mas\,yr$^{-1}$. Table\,\ref{TABLE3} and Table\,\ref{TABLE4}  give the properties 
of these five stars (sorted by Right Ascension).  

The star with $\mu$\,$=$\,$4.6$ mas\,yr$^{-1}$ is \object{2MASS J06091200-4717207} and serves as another 
demonstration of the fact that the PL relation is not always valid. This star has an amplitude 
of $\Delta m_{\rm CSS} = 1.4$ mag. 
Its magnitudes are  $m_{\rm CSS} = 18.14$ and $K_{\rm s}= 14.3$\ . The light curve is  
periodic with $P = 116$\,days. If it was a classical cool semi-regular, its period 
would put it at 125 kpc, but its proper motion would imply an excessively large tangential velocity of 
$\sim 2700$ km s$^{-1}$. Therefore, it is more probable that its distance is about ten times 
smaller. This star is underluminous with respect to the main (sequence C, i.e. fundamental mode) classical 
PL relation.

\begin{table}[!ht]
\caption[]{{\it Gaia} EDR3 parallaxes ($\pi$ in mas) and proper motions (in mas\,yr$^{-1}$)
 for our ten candidate distant stars. No correction was added to $\pi$ because of  the 0.030 mas
 zero-point of parallaxes, as recommended in the case of individual stars 
(Arenou et al.\, \cite{arenou18}; their Section 4.5).}
\label{TABLE4}
\begin{tabular}{llll}

\noalign{\smallskip}
\hline
\hline
\noalign{\smallskip}

Star    & ~~~~~~~$\pi$ & ~~~~~$\mu_{\rm RA}$ & ~~~~~$\mu_{\rm Dec}$\\

\noalign{\smallskip}
\hline
\noalign{\smallskip}
\noalign{\smallskip}

KG-19   &   $+$0.005  $\pm$ 0.021   &  $+$1.814  $\pm$ 0.024     & $-$0.367 $\pm$ 0.031\\
KG-23   &    $+$0.023 $\pm$ 0.027   &  $+$1.507 $\pm$ 0.033      & $-$0.042 $\pm$ 0.036\\
KG-72   &   $-$0.001 $\pm$ 0.033    &  $+$0.115 $\pm$ 0.035      & $-$0.994 $\pm$ 0.030\\
KG-88   &   $-$0.040 $\pm$ 0.041    &  $-$0.320 $\pm$ 0.037      & $-$1.211 $\pm$ 0.029\\
KG-93   & $-$0.018   $\pm$ 0.036    &  $+$0.461 $\pm$ 0.037      & $-$0.590 $\pm$ 0.034\\

A       &  $-$0.060   $\pm$ 0.072     &  $+$0.231 $\pm$ 0.076     & $-$0.118 $\pm$ 0.081\\
C       &  $+$0.010 $\pm$ 0.068        &  $-$0.031 $\pm$ 0.055    & $+$0.066  $\pm$ 0.056\\
D       &  $-$0.036 $\pm$ 0.057      &  $-$0.162 $\pm$ 0.058      & $-$0.238 $\pm$ 0.048\\ 
B       &  $-$0.049 $\pm$ 0.070     &  $-$0.419 $\pm$ 0.079      &  $-$0.363 $\pm$ 0.090\\
E       &  $-$0.002 $\pm$ 0.041    &  $-$0.514 $\pm$ 0.043      & $-$0.500 $\pm$ 0.039\\

\noalign{\smallskip}
\noalign{\smallskip}
\hline
\end{tabular}
\end{table}

\section{Discussion}

\subsection{Catalogue}
We provide a catalogue of the 77 LPV stars that were spectrally classified on the HES plates. They obey $\delta < -20$\degree, and  almost all obey $|b|> 30$\degree. This latitude limit is due to the HES imprint, although KG-25, KG-26, KG-40, and KG-54 are between $b=-26$\degree and $b=-29$\degree, and KG-30 is at $b=+22$\degree. 
An abridged version of the first five lines is given in Table\,\ref{TABLE5}. We list: J2000 coordinates, Catalina mean magnitude, period, CSS-band amplitude, $K_{\rm s}$(2MASS), flag, distance in kpc, uncertainty on distance, star denomination KG-nn, M or C type, and a comment. The flag is set to 1 for normal halo LPV cases. The flag is $>1$ for members of dwarf galaxies or other peculiarities. The distances of the halo LPVs are in the range 3 to 65 kpc, and their heights above or below the Galactic plane are more than 5 kpc for 57 of them.

\begin{table*}[!ht]
\caption[]{ Catalogue of Catalina southern LPVs with spectral classification and distances}
\label{TABLE5}
\begin{center}
\label{ta}
\begin{tabular}{lllrrrrrrlll}
\noalign{\smallskip}
\hline
\hline
\noalign{\smallskip}

$\alpha$ (deg)    & $\delta$ (deg) & $m_{\rm CSS}$ & $P$(d) & $\Delta m_{\rm CSS}$ & $K_{\rm s}$ & flag & $D$(kpc) & $\delta D$ & name & Spect.& Note\\

\noalign{\smallskip}
\hline
\noalign{\smallskip}
\noalign{\smallskip}

  3.1427 & $-$22.9213 & 12.93 & 171.1 & 1.93 &  9.22  & 1 &  16 & 4  & KG-01 & M-type&\\           
  4.2324 & $-$44.0113 & 14.71 & 440.1 & 2.40 &  7.07  & 1 &  13 & 4  & KG-02 & C-type&\\           
 14.9956 & $-$33.4764 & 16.23 & 196.0 & 1.07 & 12.27  & 6 &  75 & 12 & KG-03 & C-type& in Sculpt\\ 
 15.3368 & $-$33.8847 & 16.18 &  91.5 & 0.75 & 13.14  & 6 &  59 & 8  & KG-04 & M-type& in Sculpt ?\\ 
 15.4728 & $-$65.3065 & 15.71 & 162.3 & 0.50 & 12.09  & 3 &  59 & 7  & KG-05 & C-type&  C in LMC\\  

\noalign{\smallskip}
\noalign{\smallskip}
\hline
\end{tabular}
\end{center}
\end{table*}

Our list can be compared to that of Grady et al.\,(\cite{grady19}). After selecting stars with  $\delta < -20$\degree and $|b| > 30$\degree\, in the Grady list, we obtain 34 objects. Inspecting this sample shows that 8 are in the LMC, 2 are in Fornax, 1 is  in Sculptor, 2 are false 
variables\footnote{\object{CRTS J213033.1-535749} at $\alpha =322.6379$\,deg., $\delta =-53.9635$\,deg., $m_{\rm CSS}=17.65$, \KS=15.6, P(CSS)=223\,days, polluted by X Ind located at 51$''$ with P(GCVS)=225\,d; and \object{CRTS J224311.1-413113} at $\alpha = 340.7961$, $\delta = -41.5202$, $m_{\rm CSS}=16.85$, \KS=15.2, P(CSS)=255\,d, polluted by DS Gru at 46$''$ with P(GCVS)=260\,d.}. There are 14 are in common with our sample, and finally, 7 remain that are not in our list. These 7 are not within the HES plates or impossible to classify with certainty.  Inversely, of our initial 93 stars in the same sky imprint, we eliminate 16 false variable, 7 in the  LMC, 2 in Sculptor, 1 in Fornax. 
Consequently, this cleaning procedure results in samples of 21 and 67 stars from the Grady list and our selection, respectively.  

The main reason for which the Grady sample is smaller than ours is that these latter authors apply an amplitude cut, favouring high-amplitude variables that preferentially obey the sequence-C  Mira PL relation. In contrast, the stars we seek and find  are generally periodic, low-amplitude SRa/SRb variables. Grady et al. adopt the definition of \iGaia amplitude introduced by Deason et al.\,(\cite{deason17}): $A_{\rm Gaia} = \sqrt(N_{\rm obs})$$\times$$(\sigma_{FG} / FG) $ where $N_{\rm  obs}$ is the number of observations  and $FG$ is the G-band flux. Also, when building their sample, Grady et al. require $log(A_{\rm Gaia}) > -0.55$. However, this cut discards numerous interesting stars for us. To illustrate this point, we first consider all Miras and periodic SRa and SRb of the General Catalogue of Variable Stars (Samus et al.\ \cite{samus17}) and we plot the histogram of $log(A_{\rm Gaia})$ (Fig.\,\ref{fig6-HISTOAMP}). It can be seen that the large majority (73\%) of periodic SRa/SRb variables is missed with the Grady amplitude cut. 

The same conclusion is reached when we  consider  135 M-type pulsators of the Sagittarius leading arm\footnote{We select  them with $B_{\rm Sgr}$ between $-15$\degree and $+15$\degree, and $\Lambda_{\rm Sgr} < 180$\degree. They can be seen in Fig.\,4  of Mauron et al.\,(\cite{mauron19b}) (centre panel) along the sinusoid line. The definition of longitude and latitude with respect to the Sgr stream plane is given by Belokurov et al.\ (\cite{belokurov14})}.  These stars  satisfactorily obey the main PL  relation. Again, the Grady cut would eliminate more than half of these Sgr arm M-type pulsators. Consequently, we chose not to apply this \iGaia amplitude selection here, but  we use  \iGaia parallaxes and proper motions (via tangential velocities) wherever possible to eliminate stars for which the PL distance is questionable.

\subsection{Colour comparison with AGBs of globular clusters}

\begin{figure}[!ht]
\begin{center}
\includegraphics*[width=5cm, angle=-90]{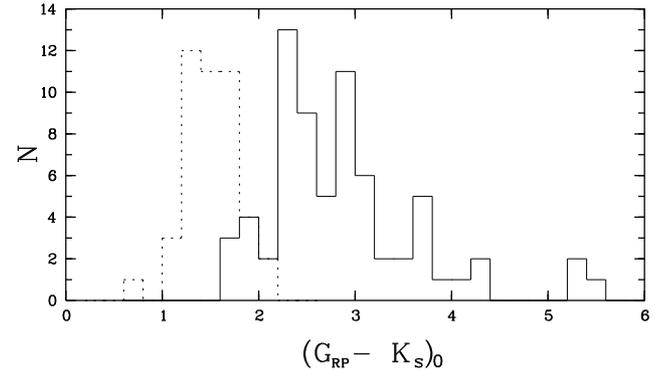}
\caption[] {  Histograms of the colour ($G_{\rm RP} - K_{\rm s}$), corrected for interstellar extinction, for  the M-type halo LPVs (continuous line), and for the LPV of metal-poor globular clusters M2, M3, M5, M13, whose metallicities are between -1.3 to -1.65 (dotted line). }
\label{fig4-HISTOGMK}
\end{center}
\end{figure}

  The colours of our halo LPVs can be compared to those of globular clusters. The {\it Gaia}-2MASS colour $G_{\rm RP} - K_{\rm s}$ is adopted here because it offers sufficient wavelength baseline, and because $G_{\rm RP}$ is the $Gaia$ filter for which amplitudes are the lowest, compared to $G$ or $G_{\rm 
BP}$. In addition $Gaia$ provides homogeneous photometry for both our LPV sample and the globular cluster stars. The interstellar extinction for the  $G_{\rm RP}$ passband is 1.83 $E(B-V)$ (Wang \& Chen\ \cite{wang19}). We first consider the four metal-poor clusters M2, M3, M5, M13 at metallicities [Fe/H] of $-1.65$, $-1.50$, $-1.29$ and $-1.53$, respectively. For these clusters, Garc\'{i}a-Hern\'{a}ndez et al.\,(\cite{garciahernandez15}) found a total of  44 AGB stars of first or second generation. The colour histograms of Fig.\,\ref{fig4-HISTOGMK} show that the LPVs studied here are almost all redder than those of these four clusters. 

\begin{figure}[!ht]
\begin{center}
\includegraphics*[width=7cm, angle=-90]{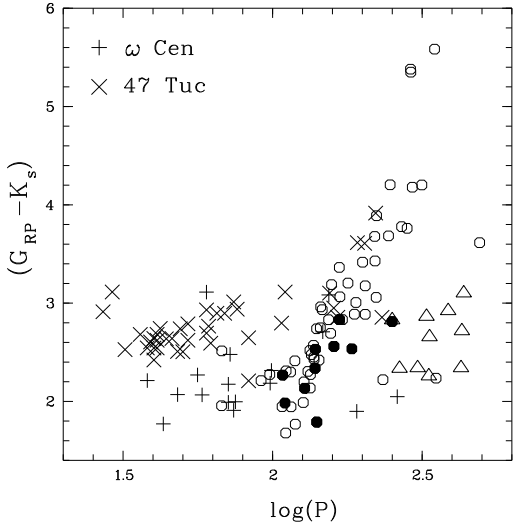}
\caption[] {  Colour ($G_{\rm RP} - K_{\rm s}$), corrected for interstellar extinction, plotted as a function of pulsation period. Circles represent the M-type LPVs of our sample. Triangles are the large-period small-amplitude group in our sample seen in Fig.\,\ref{fig2-PA}. Filled circles are the discovered distant LPVs listed in Table\,\ref{TABLE3}. The LPVs of the globular clusters $\omega$ Cen ([Fe/H]$=$\,$-1.6$) and 47 Tuc ([Fe/H]$=-0.70$) are shown for comparison, and are indicated by  plus signs and crosses, respectively. Note that a few GC LPVs with $log(P) > 2.1$ are located along the rising red main group. Uncertainties are comparable to symbol size. }
\label{fig5-PCOLOR}
\end{center}
\end{figure}

\begin{figure}[!ht]
\begin{center}
\includegraphics*[width=6cm, angle=-90]{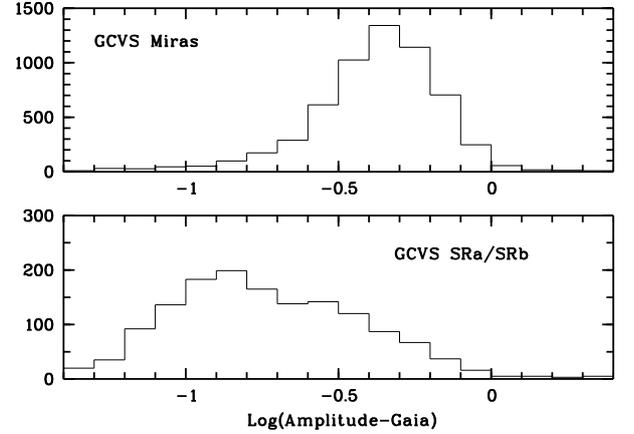}
\caption[] {Histograms of the \iGaia DR2 amplitude for Mira and SRa/SRb periodic variables of the General Catalogue of Variable Stars (Samus et al.\ \cite{samus17}). The Gaia amplitude has been defined by Deason et al.\ (\cite{deason17}) and combines the Gaia $G$ flux, its fluctuations, and the number of measurements (see Section 3.1, 2nd paragraph). The Gaia amplitude histogram is plotted in log here.}
\label{fig6-HISTOAMP}
\end{center}
\end{figure}

A complementary view is obtained by plotting colours versus periods when periods are available (Fig.\,\ref{fig5-PCOLOR}). We consider two globular cluster possessing a substancial number of identified LPVs: $\omega$ Cen with [Fe/H]\,$=-1.60$ and with 15 known red, slow variable stars 
(Lebzelter \& Wood\ \cite{lebzelter16}), and  47 Tuc ([Fe/H] $= -0.70$) with 38 variables (Lebzelter and Wood\ \cite{lebzelter05}).  Figure\,\ref{fig5-PCOLOR} shows that the majority of the cluster LPVs  have  $P<100$\,days, with a few located on a rising band with larger periods and a redder colour. We note that most of our halo LPVs, including the ten distant stars (except one), are located on this band, suggesting that many cluster-like LPVs with periods in the range  $\sim$  30 to 100 days remain to be identified in the halo.

Figure\,\ref{fig5-PCOLOR} also shows that the large-period small-amplitude LPVs seen in Fig\,2 and plotted with triangles in Fig.\,\ref{fig5-PCOLOR} are clearly separated from the other `normal' LPVs. Because they are relatively blue for their periods, we suspect that those stars do not obey the P-L relation, in a way similar to the variable V13 of 47 Tuc in its long-period mode (Fig.\,14 of Lebzelter and Wood\ \cite{lebzelter05}). Therefore, in our catalogue, distances for these stars are not given.

\subsection{Distant stars}

\begin{figure}[!ht]
\begin{center}
\includegraphics*[width=7cm, angle=-90]{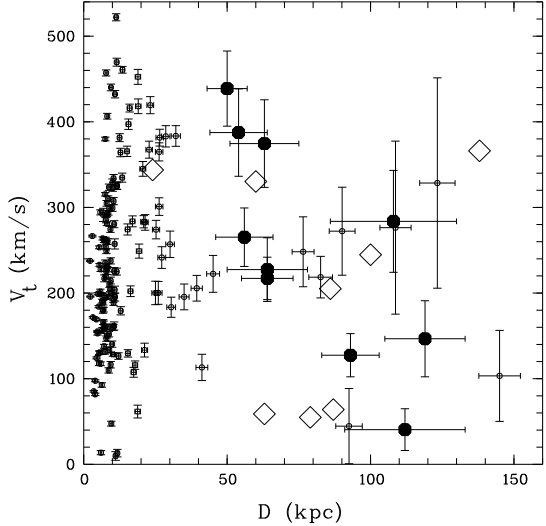}
\caption[] {Tangential velocities $versus$ heliocentric distances. No correction for the 
Sun Galactic orbital motion was applied. Our ten stars in Table\,\ref{TABLE3} are plotted with big filled circles. 
Squares are globular clusters from Harris (\cite{harris10}), with  distance 
relative uncertainty of 5\%, and proper motions from Vasiliev (\cite{vasiliev19}). 
Diamonds are  dwarf galaxies: at the top, from left to right: Sgr, Boo I, Sextans, 
Carina, and Fornax; at the bottom, Ursa Minor, Draco, and Sculptor. Distances are from 
van den Bergh (\cite{vandenbergh00}) and proper motions from Gaia DR2 (Gaia Collaboration\ \cite{gaiahelmi18b}). 
Errors for dwarf galaxies are comparable to the diamond size.}
\label{fig7-DVT}
\end{center}
\end{figure}

 In Table\,\,\ref{TABLE3} and Table\,\,\ref{TABLE4}, we identified ten cases, with amplitudes $\Delta m_{\rm CSS}$\,$>\, $0.45\,mag and $K_{\rm s}$ between 11 and 14. These objects are considered as best candidate distant AGBs, the major finding of this study. However, the applicability of the PL relation in this case remains debatable. If this is the case,  distances between 55 and 120 kpc are derived. Table\,\,\ref{TABLE4} shows that their {\it Gaia} DR2 proper motions  are typically smaller than $\sim$ 1 mas\,yr$^{-1}$. This translates to plausible tangential velocities of less than $\sim$ 450 km/s. 
In Fig.\,\ref{fig7-DVT}, we plot these ten tangential velocities as a function of distance from the Sun (filled circles). For comparison, we also plot two other families of objects: all Galactic GCs, and the few dwarf galaxies for which {\it Gaia} DR2 motions have been measured. This figure shows that our candidates have reasonable $V_{\rm t}$ velocities. Most GCs have $D< 20$ kpc. For these, the average of their velocity ($\sim$ 250 km/s) corresponds to the reflex circular velocity of the Sun, whereas the dispersion 
($\pm$\,250 km/s) reflects their intrinsic motions. The halo clusters at $D > 30$\,kpc have $V_{\rm t}$ between 0 and 400 km/s. The dwarf galaxies have a similar range. In conclusion, given the large uncertainties on the small proper motions,  the positions of our ten distant candidates  shows them to be reasonably similar to other halo populations (clusters and dwarf galaxies), supporting their halo membership.

\subsection{An underluminous AGB} 

We see above that \object{2MASS J06091200$-$4717207}  is significantly underluminous, by approximately 5 mag. The period and amplitude of this star, 116 days and 1.4 mag, respectively, are close to or larger than those of our ten candidates. To investigate the rarity of such stars, we sought similar stars in the solar neighbourhood by considering  the General Catalogue of Variable Stars (Samus et al.\ \cite{samus17}). We required {\it Gaia}  parallaxes of excellent quality, that is $\pi/\delta\pi > 10,$ and we found several underluminous cases. One example is CO Sge, an M-type Mira with $P=190$\,d and $\Delta V = 3.6$\,mag, which is confirmed by the light curve of the ASAS-SN survey (Shappee et al.\ \cite{shappee14}; Kochanek et al.\ \cite{kochanek17}). Its parallax is  $\pi = 3.45 \pm 0.18$ mas, and $D$\,$=$\,$290$\,pc.  The flux $K_{\rm s} = 4.70 \pm 0.02$  is not saturated in 2MASS. Because of variability, the mean $K_{\rm s}$-band signal is within 1.4 mag of 
this magnitude\footnote{We found a  $K_{\rm s}$ to $m_{\rm CSS}$ amplitude ratio of $\Delta K/ \Delta m_{\rm CSS} \sim 0.4$ for this kind of star 
(Mauron et al.\ \cite{mauron19b}, their Fig.\,8). A smaller ratio, $ \sim 0.15$, is given for the Johnson $V$-band for cool AGBs in Sect. 2.2 of Groenewegen\ \cite{groenewegen07}}.
 We derive from {\it Gaia} a mean K$_{\rm s}$-band absolute magnitude $M_{\rm Ks} = -2.6 \pm 1.4$, while the  SMC PL relation gives  $-6.4$. Consequently, CO Sge is  a striking example of a large-amplitude pulsator with a luminosity deficit of 3-4  mag. If our stars  are similarly underluminous, their distance $D$ and velocity $V_{\rm t}$ are smaller than proposed here.

\subsection{Contaminants}

 Here, we address qualitatively  the issue of contamination by other halo variable stars in our sample of 77 candidate halo AGB stars. For that purpose, the very recent paper by  Chen et al.\ \cite{chen20} (published when our work was being achieved) provides  very useful information. Their work is based on the Zwicky Transient Facility (Bellm et al.\ \cite{bellm19}; Graham et al.\ \cite{graham19}) scanning the entire northern sky  in three
nights, in two passbands ($g$,$r$), and to magnitude $r \sim 20.5$ ($5\sigma$); its Data Release 2 spans over 470 days.  Chen et al.\ (\cite{chen20}) analyse the DR2 light curves and provide a catalogue of $\sim 800\,000$ periodic variables. They classify  Mira and SR stars   with periods $P > 80$\,days for Miras and $P > 20$\,days for SRs, together with cuts in amplitude, with Amp($g$)~$>2.4$ for Miras and Amp($g$)~$< 2.4$ for SRs.  These latter authors find as many as 12\,000 Miras and 120\,000 SRs, in part because ZTF covers the Galactic plane. 

It is clear from their Fig.\,5, which shows luminosity versus period, that a small number of  binaries (classified ``EW" by them) approach the locus of SRs. Most  of these binaries have periods of 0.1-1days, but some are found with $P \sim 100$\,d. Their light shapes  obey characteristic relations linking the Fourier coefficients, and these relations are used to identify them (Table 1 of Chen et al.). Given that Catalina photometric quality is poorer than that of ZTF, it is very plausible that binaries contaminate the halo SR populations in which we have searched for candidates. For example, in Fig.\,11, panel (b), of  Chen et al.\ (\cite{chen20}) the top two binary ``EW" ZTF curves are quite similar to the two ZTF SRs shown at the top of their Fig.\,12, panel (a). If observed with Catalina, the curves would appear almost identical. 

 In conclusion, our halo LPVs may be contaminated by some binary variables, but this should improve in the future by exploiting the modern ZTF survey that surpasses Catalina in many aspects that are useful to separate true halo AGBs from these contaminants, particularly for small amplitudes. We also note that, for the region $\delta < -20$\degree\, not covered by ZTF, the Catalina survey has presently accumulated unique data over more than 16 years and is ongoing, and this will offer even better lightcurves.
 
\section{Summary and conclusions}

 To search for AGB stars in the outer halo ($>$ 50 kpc), we started with the CSS catalogue of LPV of Drake et al.\,(\cite{drake17}) (with $\delta < -20$\degree\,, 10000 deg$^{2}$, limiting magnitude $\sim$ 17-18, time span 8 years). 
By focussing first on the  CSS LPVs that are located within the imprint of the southern Hamburg-ESO objective prism plates ($|b| > 30$\degree), we were able to classify (M or C) 77 of them, after discarding 16 artefacts. The large majority are M-type.  Only two new, uncatalogued  C stars were found.  After excluding members of known dwarf galaxies, all  are found to be brighter than $K_{\rm s}=12.5$. This could be due to the scarcity of the sought distant objects, 
or to some instrumental limitation. 

To  search for fainter, even more  distant LPVs, we considered CSS stars in the whole Drake et al. catalogue, and required
  $K_{\rm s} > 12.5$. After discarding again false variable sources and  LPVs located in the LMC, Fornax, and Sculptor dwarf galaxies, 
we find five candidate distant stars. In our small sample of ten objects (Table\,\ref{TABLE3}),   amplitudes  $\Delta m_{\rm CSS}$ are in the range 0.45-1.5 mag, and  periods are between 100 and 230 days. We assume that the fundamental mode PL relation can provide their distance, and that they are similar to semi-regulars of the SMC. This system was chosen for its supposed low metal abundance comparable to that of our halo sample. The derived distances are found to be between $\sim$ 50 and 120 kpc; their {\it Gaia} parallaxes are not significant, and their tiny proper motions are compatible with large distances and Galactic halo membership. These ten candidates are the result of this research. In addition, a catalogue of the 77 M-type and C-type stars will be made available at the CDS.  During our analysis, we also found several cases of underluminous LPVs that deserve attention.  Finally, the halo M-type LPVs studied here ressemble the redder slow variables  of globular clusters with $P > 100$~d, but the more numerous  LPVs with smaller periods ($P \sim 30$ to $100$~d) that are found in clusters remain to be identified in the halo. 

The superb light curves and depth of the ZTF for northern stars, the data points acquired over seven additional years by the Catalina experiment for southern stars,  and the future lightcurves of the Large Synoptic Survey Telescope  can  be considered to develop this  line of research. In addition, Gaia proper motions will also improve in the near future. When follow-up spectroscopy providing radial velocities is achieved, we will hopefully have greater knowledge of  the  population of AGB stars in the outer halo of our Galaxy. 


\begin{acknowledgements}

The authors thank the anonymous referee for remarks that greatly improved the manuscript. KH acknowledges support for this work from the NASA ADAP program under NASA grant 18-ADAP18-228. NM thanks Sebastian Otero (AAVSO)  for very useful comments, Henri Reboul and Denis Puy for encouragments, and the staff of LUPM for kind assistance, especially St\'{e}phane Nou and Lydie Le Clainche. NM is grateful to the Universit\'{e} de Montpellier and the  Institut National de Physique Nucl\'{e}aire et de Physique des Particules (IN2P3; Centre National de Recherche Scientifique) for very appreciated support. This work makes use of the Catalina database (California Institute of Technology, \& NASA), the Two Micron All Sky Survey (U. of Massachusetts and IPAC/Caltech funded by NASA and NSF). We very much appreciated to use the on-line archive of  the Hamburg-ESO (Hamburg Observatory \&  European Southern Observatory) and express thanks to D. Engels. We also exploited the unique data of the European Space Agency mission {\it Gaia}. {\it Gaia} data are being processed by the {\it Gaia} Data Processing and Analysis Consortium (DPAC). Funding for the DPAC is provided by national institutions, in particular the institutions participating in the {\it Gaia} MultiLateral Agreement (MLA). The {\it Gaia} mission website is  https://www.cosmos.esa.int/gaia. The {\it Gaia} archive website is https:// archives.esac.esa.int/gaia. Finally, we used the SIMBAD/Vizier facilities offered by and operated at CDS, Strasbourg, France.

\end{acknowledgements}




\begin{thebibliography}{}

\bibitem[2019]{abia19}
Abia, C., de Laverny, P., Cristallo, S., et al. 2019, arXiv:1911.09413

\bibitem[2018]{arenou18}
Arenou, F., Luri, X., Babusiaux, C., et al. 2018, A\&A, 616, A17

\bibitem[2019]{bellm19}
Bellm, E.C., Kulkarni, S.R., Graham, M.J., et al. 2019, PASP, 131, 018002

\bibitem[2014]{belokurov14}
Belokurov, A., Koposov, S.E., Evans, N.W., et al. 2014, MNRAS, 437, 116

\bibitem[2016]{blandhawthorn16}
Bland-Hawthorn, J., \& Gerhard, O. 2016, ARAA, 54, 529

\bibitem[2014]{bochanski14}
Bochanski, J., Willman, B., West, A., et al. 2014, AJ, 147, 76

\bibitem[2018]{carlin18}
Carlin, J.L., Sheffield, A.A., Cunha, K., \& Smith, V.V. 2018, ApJL, 859, L10 

\bibitem[2020]{chen20}
Chen, X., Wang, S., Deng, L., et al. 2020, ApJSS, 249, 18

\bibitem[2004]{christlieb04}
Christlieb, N., Reimers, D., Wisotzki, L. 2004, The Messenger, 117, 40

\bibitem[2001]{clement01}
Clement, C. 2001, AJ, 122, 2587

\bibitem[2017]{clement17}
Clement, C. 2017, in {\it Wide-Field Variability Surveys: A 21$_{st}$ Century Perpective};
 EPJ Web of Conferences

\bibitem[2017]{deason17}
Deason, A.J., Belokurov, V., Erkal, D., et al. 2017, MNRAS, 467, 2636

\bibitem[2020]{deason20}
Deason, A.J., Erkal, D., Belokurov, V., et al. 2020, arXiv 2010.1380

\bibitem[2017]{drake17}
Drake, A., Djorgovski, S.G., Catelan, M., et al. 2017, MNRAS, 469, 3688


\bibitem[2018b]{gaiahelmi18b}
Gaia Collaboration, Helmi, A., van Leeuwen, F., et al. 2018b, A\&A, 616, A12

\bibitem[2015]{garciahernandez15}
Garc\'ia-Hern\'andez, D.A., Meszaros, Sz., Monelli, M., et al. 2015, ApJL, 815, L4

\bibitem[1997]{gnedin97}
Gnedin, O.Y., \& Ostriker, J.P. 1997, ApJ, 474, 223

\bibitem[2019]{grady19}
Grady, J., Belokurov, V., \& Evans, N.W. 2019, MNRAS, 483, 3022

\bibitem[2019]{graham19}
Graham, M.J., Kulkarni, S.R., Bellm, E.C., et al. 2019, PASP, 131, 78001

\bibitem[2007]{groenewegen07}
Groenewegen, M.A.T. 2006, in Why Galaxies Care About AGB Stars: Their 
importance as actors and probes, ed. F.Kerschbaum, C. Charbonnel, \& R.F. Wing, 
ASP conference series, Vol. 378, p.433 

\bibitem[2017]{harmsen17}
Harmsen, B., Monachesi, A., Bell, E.F., et al. 2016, MNRAS 466, 1491

\bibitem[2010]{harris10}
Harris, W.E. 2010, arXiv:1012.3224

\bibitem[2008]{helmi08}
Helmi, A. 2008, Astron. Astrophys. Review, 15, 145

\bibitem[2016]{hernitschek16}
Hernitschek, N., Schlafly, E.F., Sesar, B., et al. 2016, ApJ, 817, 73

\bibitem[2015]{huxor15}
Huxor, A.P., \& Grebel, E.K. 2015, MNRAS, 453, 2653

\bibitem[2016]{janesh16}
Janesh, W., Morrison, H.L., Ma, Z., et al. 2016, ApJ, 816, 80

\bibitem[2018]{jayasinghe18}
Jayasinghe, T., Kochanek, C.S., Stanek, K.Z., et al. 2018, MNRAS, 477, 3145

\bibitem[2020]{jonsson20}
J\"{o}nsson, H., Holtzman, J.A., Allende Prieto, C., et al. 2020, AJ, 160, 120

\bibitem[2017]{kochanek17}
Kochanek, C.S., Shappee, B.J., Stanek, K.Z., et al. 2017, PASP, 129, 104502


\bibitem[2005]{lebzelter05}
Lebzelter, T., \& Wood, P.R. 2005, A\&A, 441, 1117 

\bibitem[2016]{lebzelter16}
Lebzelter, T., \& Wood, P.R. 2016, A\&A, 585, A111

\bibitem[2004]{lee04}
Lee, K.H., Lee, H.M., Fahlman, G.G., \& Sung, H. 2004, ApJ, 128, 2838


\bibitem[2019a]{mauron19a}
Mauron, N., Gigoyan, K.S., Gigoyan, K.K., et al. 2019a, Astrophysics, 62, 202M  (arXiv:1901.11427)

\bibitem[2019b]{mauron19b}
Mauron, N., Maurin, L.P.A., \& Kendall, T.R. 2019b, A\&A, 626, A112

\bibitem[2012]{mcconnachie12}
McConnachie, A.W. 2012, AJ, 144, 4

\bibitem[2016]{monachesi16}
Monachesi, A., Bell, E.F., Radburn-Smith, D.J., et al. 2016, MNRAS, 457, 1419

\bibitem[2001]{odenkirchen01}
Odenkirchen, M., Grebel, E.K., Rockosi, C.M., et al. 2001, ApJ, 548, L165


\bibitem[2017]{samus17}
Samus, N.N., Kazarovets, E.V., Durlevitch, O.V., et al. 2017, Astronomy Reports, 61, 80

\bibitem[2011]{schlafly11}
Schlafly, E.F., \& Finkbeiner, D.P. 2011, ApJ, 737, 103

\bibitem[2017]{sesar17}
Sesar, B., Hernischek, N., Mitrovi\'{c}, S., et al. 2017, AJ, 153, 204

\bibitem[2014]{shappee14}
Shappee, B.J., Prieto, J.L., Grupe, D., et al. 2014, ApJ, 788, 48


\bibitem[2007]{soszynski07}
Soszy\'{n}ski, I., Dziembowski, W.A., Udalski, A., et al. 2007, Acta Astronomica, 57, 201

\bibitem[2019]{starkenburg19}
Starkenburg, E., Youakim, K., Martin, N., et al. 2019, MNRAS, 490, 5757

\bibitem[2020]{stringer20}
Stringer, K.M., Drlica-Wagner, A., Macri, L., et al. 2020, arXiv:2011.13930

\bibitem[2018]{thomas18}
Thomas, G.F., McConnachie, A.W., Ibata, R., et al. 2018, MNRAS, 481, 5223


\bibitem[2000]{vandenbergh00}
van den Bergh, S. 2000, {\it The Galaxies of the Local Group}, Cambridge University Press

\bibitem[2019]{vasiliev19}
Vasiliev, E. 2019, MNRAS, 484, 2832

\bibitem[2019]{wang19}
Wang, S., \& Chen, X. 2019, ApJ, 877, 116

\bibitem[2009]{whitelock09}
Whitelock, P.A., Menzies, J.W., Feast, M.W., et al. 2009, MNRAS, 394, 795

\bibitem[2015]{xue15}
Xue, X.-X., Rix, H.-W., Ma, Z., et al. 2015, ApJ, 809, 144

\bibitem[2019]{yang19}
Yang, C., Xue, X., Li, J., et al. 2019, ApJ, 880, 65


\end{thebibliography}
\end{document}